# Lattice complexity and fine graining of symbolic sequence

**Da-Guan Ke *, Hong Zhang, Qin-Ye Tong**

(Department of Biomedical Engineering, Zhejiang University, Hangzhou 310027,China)

**Abstract**

A new complexity measure named as *Lattice Complexity* is presented for finite symbolic sequences. This measure is based on the symbolic dynamics of one-dimensional iterative maps and Lempel-Ziv Complexity. To make *Lattice Complexity* distinguishable from Lempel-Ziv Complexity, an approach called *fine-graining process* is also proposed. When the control parameter *fine-graining order* is small enough, the two measures are almost equal. While the *order* increases, the difference between the two measures becomes more and more significant. Applying *Lattice Complexity* to logistic map with a proper order, we find that the sequences that are regarded as complex are roughly at the edges of chaotic regions. Further derived properties of the two measures concerning the fine-graining process are also discussed.

## 1 Introduction

Researchers from around the world are studying on characterizing "complexity". However, despite the miscellaneous literatures on this topic, no one can answer the primordial problem: What is complexity? The reason may lie in the nonexistence of an absolute character of complexity. And many researchers agree that such a character may be impossible to be found.

Currently, there are numerous "definitions" of complexity, each of which has its own benefits and shortages. Since concrete "definitions" seem make no help to understand the problem entirely, we should look into some paths leading from simplicity to complexity to gain an enlightened perspective.

Bai-Lin Hao had enumerated some approaches to complex phenomena [1]:

(1) Using simple rules repeatedly might produce extremely complex behaviors or patterns.

(2) Projecting a high-dimensional space onto a low-dimensional space may make a physical process originally in the former space seem more complex.

(3) Wrong reference systems may bring unnecessary complexity.

On the other hand, complexity characters are usually based on the symbolizations using particular coarse graining methods to describe systems' behavior. The symbolizations should distinguish different hierarchies. For example, the six lowercase letters u, d, c, s, b and t represent six types of quarks, and the three letters p, n and e represent proton, neutron and electronic respectively. The three kinds of subatomic particles compose atoms. Four sorts of nucleotides that form DNA molecules are denoted by a, t, c and g. All living creatures on the earth have their own DNA consisting of such nucleotides. This hierarchical problem is associated to the scale problem. On one scale a sequence may be thought complex, while on another scale it should be considered simple and even be represented by only one symbol. In other words, the complexity problem may be a relative problem that should be discussed in different hierarchies.

* Email address: kdg@zju.edu.cn

Here arises a natural question: Is it possible to reflect the relativity and the hierarchy in a complexity measure?

For simplicity, we limit ourselves to the complexity measurement of every single symbolic sequence. Since Kolmogorov proposed his famous concept of complexity [2], some estimation algorithms have been presented. Among them, Lempel-Ziv Complexity, sometimes wrongly equated with *Kolmogorov Complexity*, was thought the most “elegant” [3]. Lempel-Ziv Complexity has been used on spatiotemporal patterns [4], brain data [5] and speech sound [6] etc. However, many people including Lempel and Ziv [7,8] may agree that because Lempel-Ziv Complexity only use two kinds of simple operations, replication and insertion, to simulate the general Turing machine defined in *Kolmogorov Complexity*, the algorithm ineluctably has some defects as follows:

First, what Lempel-Ziv Complexit*y* measures is only randomness, whereas complexity may be different with randomness, as Grassberger has shown with three pictures in Ref. [9]. An ideal complexity measure is generally thought to have the relationship with randomness as shown in fig.1.

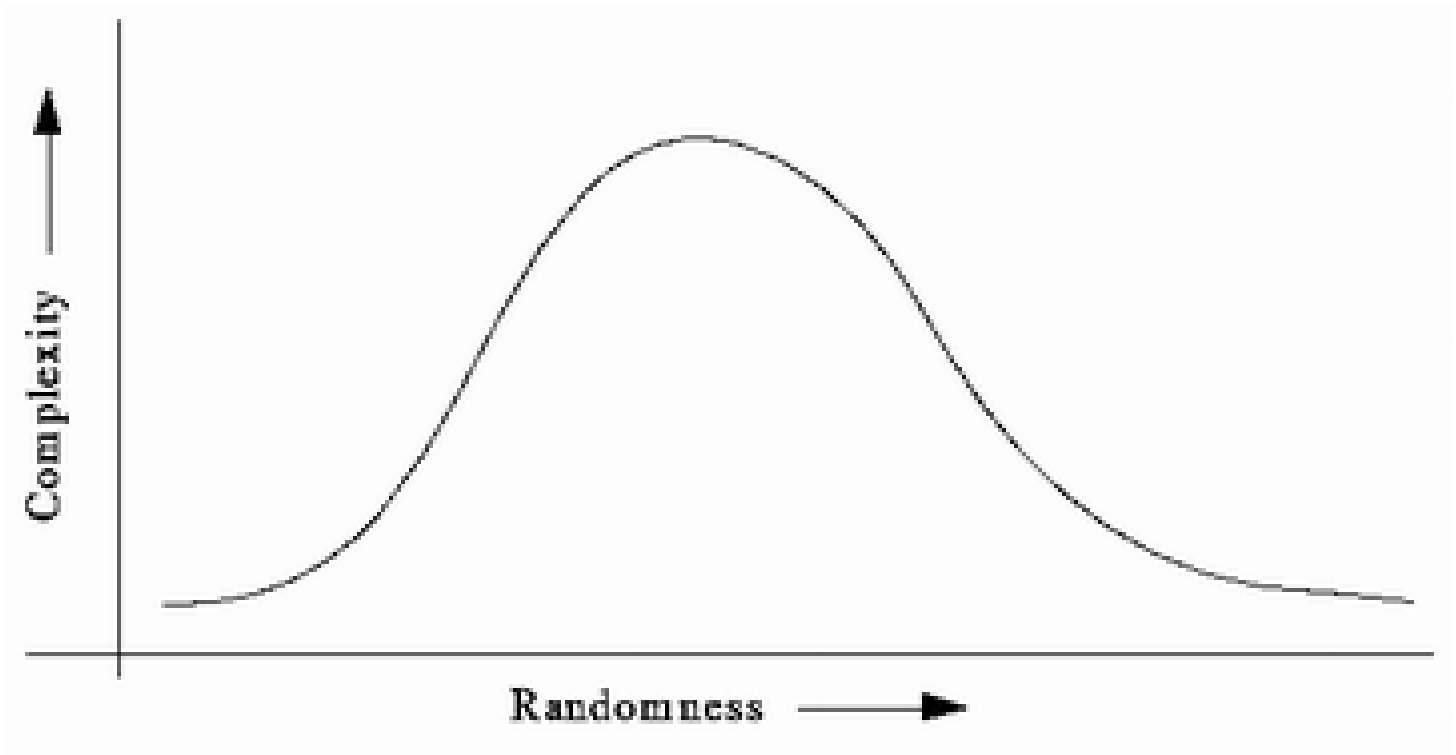

Fig. 1 Relationship between ideal complexity and randomness

Second, by defining a general Turing machine, *Kolmogorov Complexity* regards the length of the shortest program that generates a given symbolic sequence as the complexity measure of that sequence. By using a simplified computation model, Lempel-Ziv Complexity counts any chaotic pseudo-random sequence built by iterative maps as the most complex case. But since the generating program of such a sequence is very short, it is not appropriate to think the *Kolmogorov Complexity* of the sequence is high.

Third, some artificial sequences with evident regularity, de Bruijn sequences for instance, are reckoned as complex sequences by *Lempel-Ziv complexity*. It is also a serious flaw.

Obviously, a new complexity measure that preserves the advantages of Lempel-Ziv Complexity but modifies its defects will bring some progress in the complexity study. A new question is: Can we design a machine more complicated than that of Lempel and Ziv to achieve this purpose?

In this paper, we do not attempt to build an overall structure, which was disapproved by Anderson [10], but want to try to answer the above question.

# 2 Lattice Complexity and Lempel-Ziv Complexity

## 2.1 Lattice Complexity

As we know, many iterative maps including interval maps and circle maps, such as those shown in Fig.2, may take complex dynamical properties and cause chaos. Chaotic orbits can be dense everywhere in the spaces of the systems. At the same time, all the orbits are presented by symbolic sequences corresponding with some kind of coarse graining of the spaces [11,12]. A common coarse-graining process usually takes $i$ critical points as the thresholds to partition the whole space into ($i$ + 1) segments denoted by ($i$ + 1) different symbols. That will inevitably lose many details. But with the degree of the coarse graining becoming lower, the size of every segment becomes smaller.

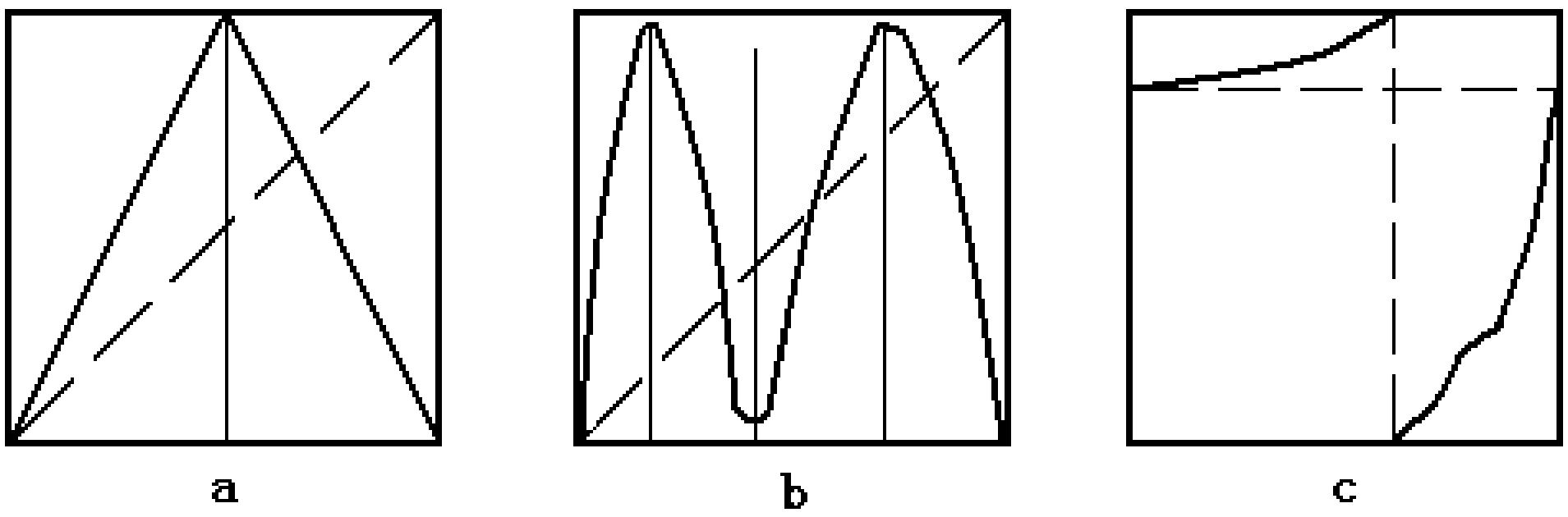

Fig. 2 (a) Unimodal map, (b) multimodal map, (c) circle map.

With respect to their modes of iterative mapping, all the points of orbits in nonlinear systems can be divided into five classes, i.e. the fixed, the periodic, the recurrent, the non-wandering and the wandering [13]. But when the coarse graining reaches to a certain degree, we need only considering the following two types, the recurrent and the non-recurrent. If we suppose each symbol only denote a neighborhood of exactly one point, within a given symbolic sequence some sub-sequences can naturally be connected to some particular trajectories of particular kinds of orbits. Such sub-sequences will be viewed as simple objects and be named as iterative sequences, including the periodic sequences with recurrent points (but with no repeated symbol within one period) and the chaotic sequences with non-recurrent points.

Combining the idea of symbolic dynamics and the algorithm of Lempel and Ziv, we design a new kind of complexity measure—*Lattice Complexity*. The complexity of any symbolic sequence is quantified as the number of the *lattices* in the sequence.

A *lattice* is defined as a sub-sequence with the properties as follows:

1) A lattice includes an iterative sequence as its prefix.
2) A lattice can remember all the exhaustive history of the sequence and can simply repeat any series of successive operations in the memory, as the replication function defined by Lempel and Ziv.

3) The last symbol of a lattice must be created by the insertion unless the end of the whole sequence has been reached.

Let us introduce the detail of the procedure.

Suppose there is a sequence $s_1 s_2 \ldots s_n$. Let $LC(n)$ denote the *Lattice Complexity* of the sequence of length $n$. We may consider now the step in the midway, at which a prefix $s_1 s_2 \ldots s_{r-1}$ ($r < n$) has already been built with an inserted symbol $s_r$, represented as

$$s_1 s_2 \cdots s_n \to s_1 s_2 \cdots s_{r-1} s_r \vee s_{r+1} \cdots .$$

The mark "$\vee$" following $s_r$ means that $s_r$ is inserted as the last symbol of a complete lattice. Another lattice is starting with $s_{r+1}$. Let us give the further proceedings.

Step 1: In this step we assume the iterative sequence is a chaotic sequence and turn to the next step when we find the sequence become a periodic sequence. Let $Q = s_{r+1}$. We do not make any operation because there is only one symbol. Then let $Q = s_{r+1} s_{r+2}$. Supposing the last letter of $Q$ is $x$, at this time we see $x = s_{r+2}$. Now we look into whether or not $x$ is equal to any letter of $Q\pi$ (here $\pi$ means to exclude the last letter of the anterior word so that we get $Q\pi = s_{r+1}$). If not, we recursively let $Q = s_{r+1} s_{r+2} s_{r+3}$ and $x = s_{r+3}$, and check whether or not $x$ is equal to any letter of $Q\pi$. So on and so forth until we find $x$ repeats one of the letters of $Q\pi$ and then turn to the next step.

Step 2: Suppose $x = s_{r+j}$ and its equal symbol in $Q\pi$ is $s_{r+i}$ ($0 < i < j$). Let $P = s_{r+i+1}$ and let $R = s_{r+j+1}$. We observe whether or not $P$ is equal to $R$. If it is true, then let $P = s_{r+i+1} s_{r+i+2}$ and let $R = s_{r+j+1} s_{r+j+2}$ to see whether or not $P$ is equal to $R$. If it is true, let $P = s_{r+i+1} s_{r+i+2} s_{r+i+3}$ and let $R = s_{r+j+1} s_{r+j+2} s_{r+j+3}$. Repeating this procedure until $P \neq R$, we obtain a complete iterative sequence $QR\pi$. Go to the next step.

Step 3: Let $Q = QR$. Suppose the last letter of $Q$ is $s_{r+k}$. If $Q$ can be regarded as a copy of some segment of $sQ\pi$, where $s = s_1 s_2 \ldots s_r$, then let $Q = s_{r+1} s_{r+2} \ldots s_{r+k} s_{r+k+1}$. If $Q$ can still be created by copying a sub-sequence of $sQ\pi$, let $Q = s_{r+1} s_{r+2} \ldots s_{r+k} s_{r+k+2}$; continue in this manner until you reach the end or you find that $Q$ can not be regarded as a replication of any segment of $sQ\pi$. In the latter situation, we have to make an insertion to create the last letter of $Q$ and append the mark "$\vee$" to this letter to separate the current *lattice* with the next one, which just starts by the null-sequence $\Lambda$. Thus we return to the state before Step1.

It can be seen that the marks of insertion divide the whole sequence into a particular number of *lattices*. If the sequence ends without mark "$\vee$", we say that the last *lattice* is incomplete. If the last symbol of a sequence is created by insertion, we say the second to last lattice of this sequence is complete and the last (and certainly incomplete) *lattice* is starting from the null-sequence $\Lambda$. In either of these two cases, the number of lattice is equal to the number of marks "$\vee$" plus one.

Assuming that every point is represented by a distinct symbol, a sequence made by any iterative process should be either the chaotic sequence (without repeated symbols) or the periodic sequence (with repeated symbols in strictly periodic manner). Since no point of a determinate map can simultaneously be both the non-recurrent point (that creating chaotic orbits) and the recurrent point (that creating periodic orbits), once a chaotic sequence has a repeated symbol as its last symbol, the iterative sequence becomes a periodic sequence. And when we find a symbol that does not obey the periodic rule, we should think the iterative map is completed. But the *lattice* can still extend if we find that the iterative sequence with a certain following sub-sequence together as a whole has already appeared in the history including the sub-sequence itself except its last letter.

After a lattice has been completed, a new iterative sequence (as a prefix of the next *lattice*) appears.

## 2.2 Relationship between Lattice Complexity and Lempel-Ziv Complexity

From the above-mentioned algorithm of *Lattice Complexity*, it is easy to see that if we omit the first two steps and directly execute Step 3 with $Q = s_{r+1}$ and $R = \Lambda$, the procedure becomes exactly that of Lempel-Ziv Complexity. The characters making *Lattice Complexity* different with Lempel-Ziv Complexity are that every lattice has an iterative sequence as its prefix, and that only after the iterative sequence has been completed we begin to check whether or not the lattice can be made by copy.

Now, using $LZC(n)$ to denote Lempel-Ziv Complexity and using dot "." to denote the insertion operation in the calculation of Lempel-Ziv Complexity, we may investigate the relationship between $LC(n)$ and $LZC(n)$.

For example, suppose we have a binary sequence $s = 1001100001110000011110110$1. The calculations of Lempel-Ziv Complexity and *Lattice Complexity* are shown as follows:

Lempel-Ziv Complexity: $1 \cdot 0 \cdot 01 \cdot 1000 \cdot 0111 \cdot 00000 \cdot 1111 \cdot 01101 \cdot$

*Lattice Complexity*: $1001 \vee 100001 \vee 1100000 \vee 11110 \vee 1101 \vee$

Here we get $LZC(s) = 8$ and $LC(s) = 6$.

First several steps of these calculations are explained as the following.

According to Lempel-Ziv Complexity, every sequence starts from the null-sequence $\Lambda$ so that the first two digits $10$ are inserted as new symbols. The third digit $0$ can be made by replication. But the word $01$ cannot be copied from the history $100$, meaning that an insertion operation has to be made. The following sub-string $100$ can be created by replication, but $1000$ has no prototype in the history $1001100$ so that we should insert the eight symbol. The rest part of the sequence is processed in the similar way.

According to *Lattice Complexity*, the first two digits $10$ are regarded as a part of an iterative sequence, which may be a chaotic sequence because every symbol is distinct, and then we need not make any insertion. The third digit is $0$, indicating that the current iterative sequence is a sequence of fixed point (a special periodic sequence). The forth digit $1$ do not continue to be the fixed point 0 so that it is an inserted point marking off the first lattice. The following five digits $10000$ make the second iterative sequence, and the digit $1$ next to it means the iterative sequence has been completed. Since the sub-sequence $100001$ cannot be produced by replication, the last digit $1$ is an insertion and we find the second complete lattice. In the same manner we find other lattices.

Insertion operations in Lempel-Ziv Complexity can usually be used to demarcate *lattices* in *Lattice Complexity* except the following two cases:

1) There is a sub-sequence without any repeated symbol, as the first two digits in the above example;
2) After an insertion operation has been made, a new periodic sequence is longer than the same periodic sequence appeared before, as the eighth digit in the above example.

Obviously, Lempel-Ziv Complexity is always greater than or equal to *Lattice Complexity*.

That is:

$$1 \le LZ(n) \le LZC(n)\,. \tag{1}$$

Supposing the difference caused by the first case is $\Delta 1(n)$, and the difference value caused by the second case is $\Delta 2(n)$, we have the whole difference as

$$LZC(n) - LZ(n) = \Delta 1(n) + \Delta 2(n)\,. \tag{2}$$

Assuming the symbol set (the alphabet) $S$ contains $\alpha$ symbols and every symbol occurs with the same possibility $\alpha^{-1}$, we see that any particular sequence of length $n$ occurs with the possibility $\alpha^{-n}$. When $n$ yields infinite, Lempel and Ziv have proven [7] that the Lempel-Ziv Complexity of almost all sequences will be close to the upper bound. Namely, almost all sequences should be thought complex. We will show that to a large extent Lattice Complexity does not negate this result.

In the first case, within every lattice there can be at most $\alpha$ insertions of Lempel-Ziv Complexity. The number of such lattices is finite, because there are $\alpha!$ permutations of the $\alpha$ symbols. We obtain:

$$\lim_{n\to\infty} \Delta 1(n) \le \alpha! \tag{3}$$

In the second case, since it has been already discovered that $\Delta 1(n)$ is associated with $\alpha$, for convenience we may assume that $n$ is a sufficiently large constant and then investigate $\Delta 2(\alpha)$ lonely.

Let $k$ be the length of the second repeated periodic sub-string whose beginning digit is the same as the beginning digit of the first iteration (within the first periodic sequence). The period $p$ of any periodic sequence is certainly less than ($k-1$), then for $k \ge \alpha + 1$ a periodic sequence of length $k$ takes the possibility of less than $\alpha! \cdot \alpha^{-k}$ to appear, and for $k < \alpha + 1$, less than $\alpha! / (\alpha - p + 1)! \cdot \alpha^{-k}$. If there is such an iterative sequence followed (not immediately) by a same iterative sequence of $(k+1)$ successive repeated symbols (with the possibility $\alpha^{-(k+1)}$), we will find that the second case happen first time and $\Delta 2(\alpha) = 1$. If later the same iterative sequence appears again with $(k+2)$ repeated symbols (with the possibility $\alpha^{-(k+2)}$), we get $\Delta 2(\alpha) = 2$. When the period $p$ is 1, the minimum $k$ is 2. Supposing $m$ is a finite positive integral smaller than $n$, we have

$$\Pr[\Delta 2(\alpha) \ge 1 \mid p = 1] = \alpha \cdot \alpha^{-k} \cdot \alpha^{-(k+1)} = \alpha^{-2k} \le \alpha^{-4};$$

$$\Pr[\Delta 2(\alpha) \ge 2 \mid p = 1] = \alpha \cdot \alpha^{-k-(k+1)-(k+2)} = \alpha^{-3k-2} \le \alpha^{-8};$$

$$\ldots$$

$$\begin{aligned} &\Pr[\Delta 2(\alpha) \ge m \mid p = 1] \\ &= \alpha \cdot \alpha^{-k-(k+1)-(k+2)\cdots-(k+m)} \\ &\le \alpha^{1-2-(2+1)-(2+2)\cdots-(2+m)} \\ &= \alpha^{1-\frac{1}{2}(m+1)(m+4)} \end{aligned} \tag{4}$$

More generally, for any $p = i$, we have

$$\begin{aligned}
&\Pr[\Delta 2(\alpha) \geq m \mid p = i] \\
&= \frac{\alpha!}{(\alpha - i + 1)!} \cdot \alpha^{-k-(k+1)-(k+2)\cdots-(k+m)} \\
&\leq \frac{\alpha!}{(\alpha - i + 1)!} \cdot \alpha^{-(i+1)-(i+1+1)-(i+1+2)\cdots-(i+1+m)} \\
&= \frac{\alpha!}{(\alpha - i + 1)!} \cdot \alpha^{-\frac{1}{2}(m+1)(m+2i+2)}
\end{aligned} \tag{5}$$

When $\Delta 2(\alpha) = m$,

$$p \leq k - 1 < \frac{n - m + 1}{m} < \frac{n}{m}. \tag{6}$$

We see that

$$\begin{aligned}
&\lim_{\alpha \to \infty} \Pr[\Delta 2(\alpha) \geq m] \\
&< \lim_{\alpha \to \infty} \{\sum_{i=1}^{n/m} \Pr[\Delta 2(\alpha) \mid p = i]\}, \quad (1 \leq m < n). \\
&= 0
\end{aligned} \tag{7}$$

It is clear that when $\alpha$ approaches infinity, $\Delta 2(\alpha)$ converges to with probability 1. In fact, from (4) and (5), we may also see that even if the number $\alpha$ is very small, for most sequences of length $n$, $\Delta 2(\alpha)$ is still not more than a positive integral $m$. Of course we can construct some exceptions. But the constructed sequences have such a degree of regularity that we should hope that *Lattice Complexity* and Lempel-Ziv Complexity give entirely different results.

Combining the two cases, we have:

$$LZC(\alpha, n) - LC(\alpha, n) \leq \alpha! + m \tag{8}$$

For a fixed $\alpha$, $\alpha! + m$ is generally a finite constant. This matches the property of Kolmogorov Complexity that the difference caused by different Turing machines should not be more than a finite constant. For this reason, like Lempel-Ziv Complexity, Lattice Complexity also approaches the upper bound $LC(n) \sim \frac{n}{\log_\alpha n}$ when $n \to \infty$. Its normalized form is

$$LCI_n = \frac{LC(n) \log_\alpha n}{hn}, \tag{9}$$

Where, $h$ is the information entropy to the base $\alpha$ [7].

But this normalized measure of *Lattice Complexity* will lose its meaning when $\alpha$ enlarges as quickly as $n$.

## 2.3Complexity of the Symbolic Sequence of Chaotic Orbit

It has already been known that the difference between Lempel-Ziv Complexity and *Lattice Complexity* depends on the size $\alpha$ of the symbol set $S$. When we use these two kinds of complexity measures to measure the complexity of symbolic sequences, the results certainly depend on the parameter $\alpha$. So $LZC(\alpha,n)$ and $LC(\alpha,n)$ should be proper expressions of these two measures.

Now, with these two kinds of measures, let us investigate the complexity of chaotic orbits of various iterative maps as shown in Fig. 2. By chaotic orbit we mean the non-periodic orbit that can be dense everywhere in the space of the system, according to the Devaney's definition of chaos [14].

For example, to characterize the tent map as shown in Fig. 2(a), we usually divide the interval into two partitions with the single critical point, therefore $\alpha = 2$. The degree of coarse graining is the highest. With a sufficiently large $n$, we have $\Delta 1(2,n) \le 2$. Such difference is trivial. But when the degree of coarse graining becomes lower and the parameter $\alpha$ increases, the upper bound of the difference $\Delta 1$ also augments. If $\alpha$ grows as $O(\alpha^2)$, $\Delta 1(\alpha,n)$ will grows as $O((\alpha^2)!)$. When $\alpha$ yield infinity, the width of each sub-interval represented by one symbol yield infinitesimal. Because any non-recurrent point of the system is within one particular sub-interval, and because the points of a chaotic orbit are non-recurrent and dense everywhere, when $\alpha$ grows as quickly as $n$, every symbolic sequence will become a single chaotic sequence. Then we have:

$$\lim_{\alpha\to\infty}\lim_{n\to\infty} LC(\alpha,n) = 1 \qquad (10)$$

The above-mentioned equation (10) is also valid for periodic orbits and quasi-periodic orbits, but the necessary and sufficient condition is easier to be satisfied than that for chaotic orbits. Later in section 5, we will give more discussions. Other proofs of some derived propositions can also be found in section 5.

Now, we have already achieved the main purpose: By using *Lattice Complexity*, chaotic orbits can be regarded as simple as periodic orbits. But by using Lempel-Ziv Complexity, because appearing a new symbol means an insertion operation, when $\alpha \sim n$ and $\alpha \to \infty$, we have $LC(\alpha,n) \sim \alpha$, meaning that chaotic orbits can only be regarded as complex.

# 3 The Fine-graining Method for Complexity Measure

The difference between *Lattice Complexity* and Lempel-Ziv Complexity depends on $\alpha$ while $\alpha$ depends on the partition of the system. Concerning the partition, there are two kinds of commonly used methods, Homogenous Partition and Generating Partition [15], denoted by $P^H$ and $P^G$ respectively. The former divides the whole interval into $N^H$ equal sub-intervals. The latter makes boundaries of the $N^G$ sub-intervals that always map to themselves, describing effectively the dynamical characteristic of the system. In Fig.3, we show these two kinds of partitions by a simple uni-modal map.

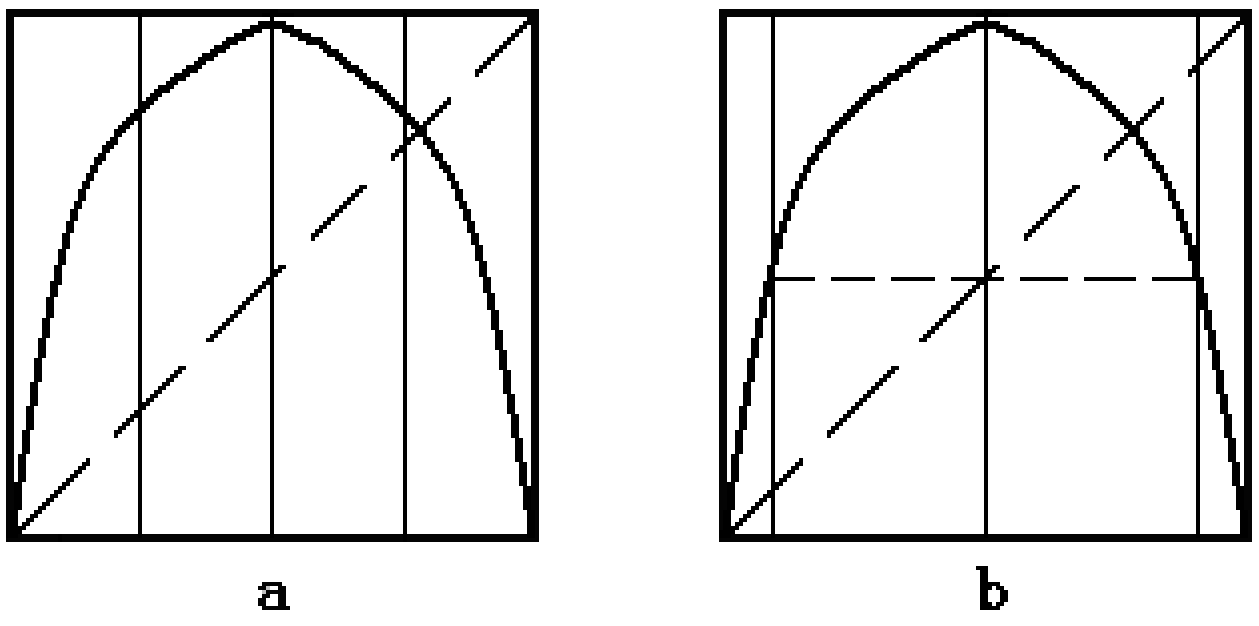

Fig. 3 (a) Homogenous Partition (b) Generating Partition

No matter what kind of partition is employed, denoting every sub-interval by a distinct symbol and taking a point $x$ as the starting point of an orbit, we have a symbolic sequence $s(x)$ of the orbit. Complexity measures are used to quantitatively characterize the complexity of $s(x)$. As the partitioned sub-intervals become finer, the alphabet becomes larger, and the behavior of *Lattice Complexity* becomes more different from that of Lempel-Ziv Complexity.

For the time series with much randomness (brain data for example), we may use Homogenous Partition. While for the data of systems whose dynamical characters have already been known, we should take Generating Partition. Both these two partitions can be further refined. But how to refine an existing symbolic sequence when we know nothing about the partition?

Suppose there is a symbolic sequence $s = s_1 s_2 \cdots s_n$ of length $n$, the alphabet $S$ related to the original partition has $\alpha$ letters. If we regard every word of $r$ letters as a new symbol, by shifting from the beginning to the end of the sequence, we will get a new sequence $s_r$ of length $(n - r + 1)$. The new alphabet $S^r$ of all possible symbols will have the size of $\alpha^r$. We call $r$ the fine-graining order of symbolic sequence to the base $\alpha$.

Symbolic dynamics give the relationship among the orbit points, symbols and the reverse functions of the maps. Then fine-graining order is identical to the refining order of the generating partition. One shift is related to one step of the iterative map function.

We call the set $S$ the *basic alphabet* and the set $S^r$ of all possible refined symbols the *fine-graining alphabet*. With the fine-graining order increasing, the number of all possible refined symbols will grow exponentially. It is impossible for a sequence of finite length to contain all these symbols. So there exist a map from $S^r$ to the *real alphabet* $S_r^*$ containing all the symbols really appear. We may use $\beta$ to denote the size of the *real alphabet* $S_r^*$ and express the map as

$$f: \qquad S^r \to S_r^*, \quad \alpha^r \to \beta .$$

In real applications, previous knowledge, e.g. the classification of amino acids, can be represented in this map.

The methods mentioned above can be used to measure the complexity of the symbolic sequence obtained by a certain coarse-graining technique. Different fine-graining orders represent different “scales” in symbolic sequence. It may answer the question arising from the so-called over-coarse-graining effect [16,17] in complications. Note that, there is a widespread misunderstanding in many literatures that Lempel-Ziv Complexity is only applicable to binary string. Actually, the fine-graining method can freely be applied in both *Lattice Complexity* and Lempel-Ziv Complexity. There is also no need to fix the size of the basic alphabet to $2$. In the normalized form of our measure as (9), the base of the logarithms is $\beta$.

When $\alpha$ has already been known, it is convenient for Lempel-Ziv Complexity and *Lattice Complexity* to take fine-graining order $r$ as a parameter to replace $\alpha$. We should denote the two measures of the sequence $s$ by $LZC(r,n)$ and $LC(r,n)$ respectively. Such expressions will be employed in the subsequent sections.

# 4 Application on One-dimensional Chaotic System

As we known, for a Logistic map

$$x_{n+1} = 1 - \mu x_n\,,\quad \mu \in [0,2]\,,$$

if we take one point within the range of $(-1,1)$ as a beginning of this iterative mapping, we will get sequences of different complex with different $\mu$. When $\mu$ increases to $\mu^* = 1.40115515\cdots$, the system enters into the chaotic region with periodic windows.

By letting $\mu$ go from $3.5$ to $4.0$ with the increment $\Delta\mu = 0.0001$, we took $5001$ symbolic sequences of length $8204$ from $5001$ iterative trajectories. Each trajectory has its first $25000$ points deleted as transient. The complexity measures of the sequences are shown in the following figures.

As shown in Fig. 4, *Lattice Complexity* is similar to Lempel-Ziv Complexity when $r = 1$. But the difference enlarges with the increasing of fine-graining order (Fig. 5 and Fig. 6). Any way, the two measures share some similar characteristics: They both reach to stable values when the order $r$ increases to a certain point; the more randomness the sequences have, the easier the measures obtain stable values. Just before converging to the stable states, the results of both measures are in the order with respect to the randomness of the sequences, one from least to greatest and another in opposite direction.

There is a minimum fine-graining order that makes Lempel-Ziv Complexity and *Lattice Complexity* of a random or pseudo-random sequence obtain stable values. We call such an order the *critical order* of the sequence and denote it by $r^*$. It is found that the sequences have less randomness always have larger critical order. Taking the minimum critical order among those shown in Fig. 5 and Fig. 6, i.e. $r=29$, Lempel-Ziv Complexity and *Lattice Complexity* behave in contrary manners as shown in Fig. 7 and Fig. 8.

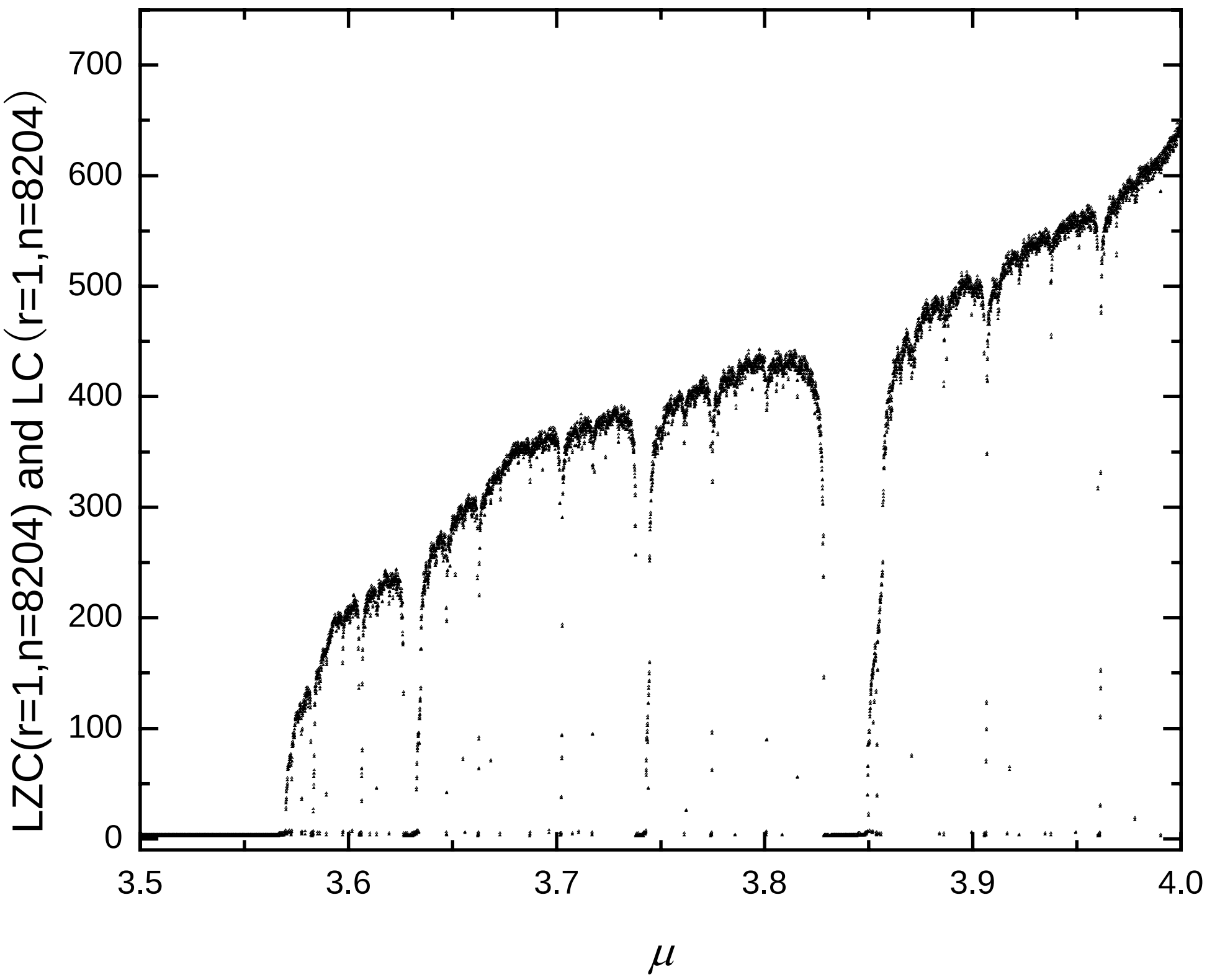

Fig. 4 *Lattice Complexity* and Lempel-Ziv complexity of fine-graining order $r = 1$

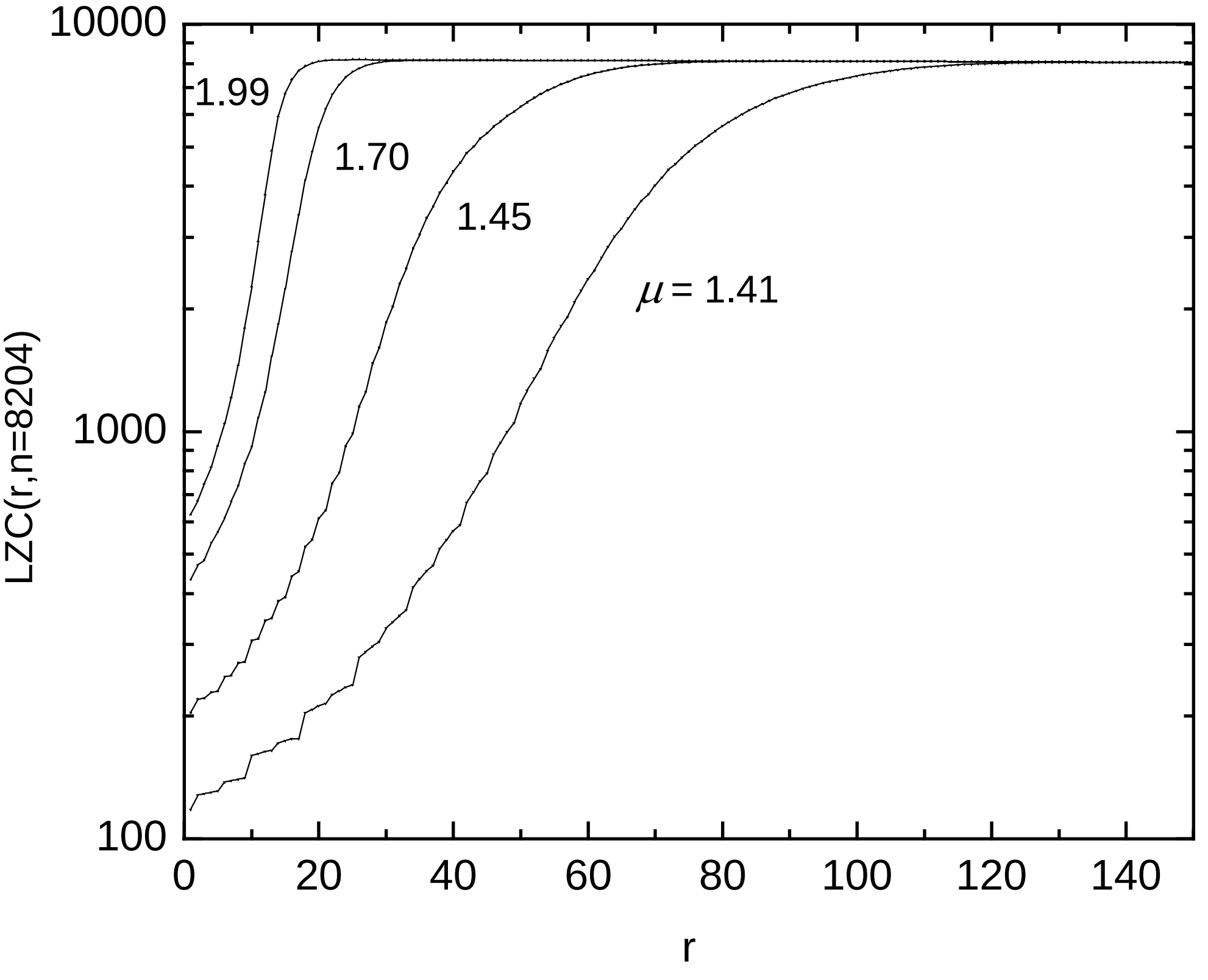

Fig. 5 Fine-graining spectrum of Lempel-Ziv Complexity for some chaotic sequences

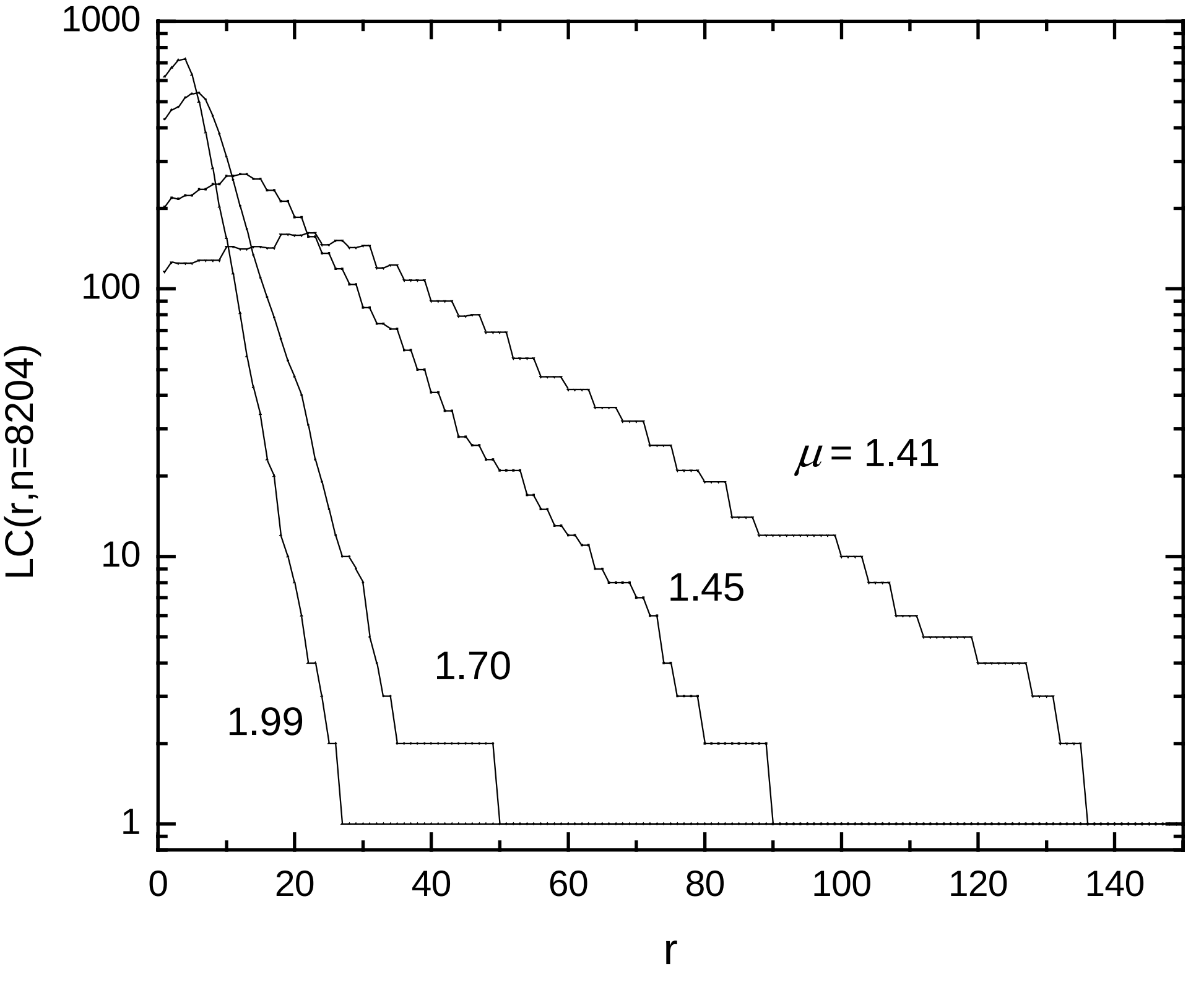

Fig. 6 Fine-graining spectrum of *Latttice Complexitie* for some chaotic sequences

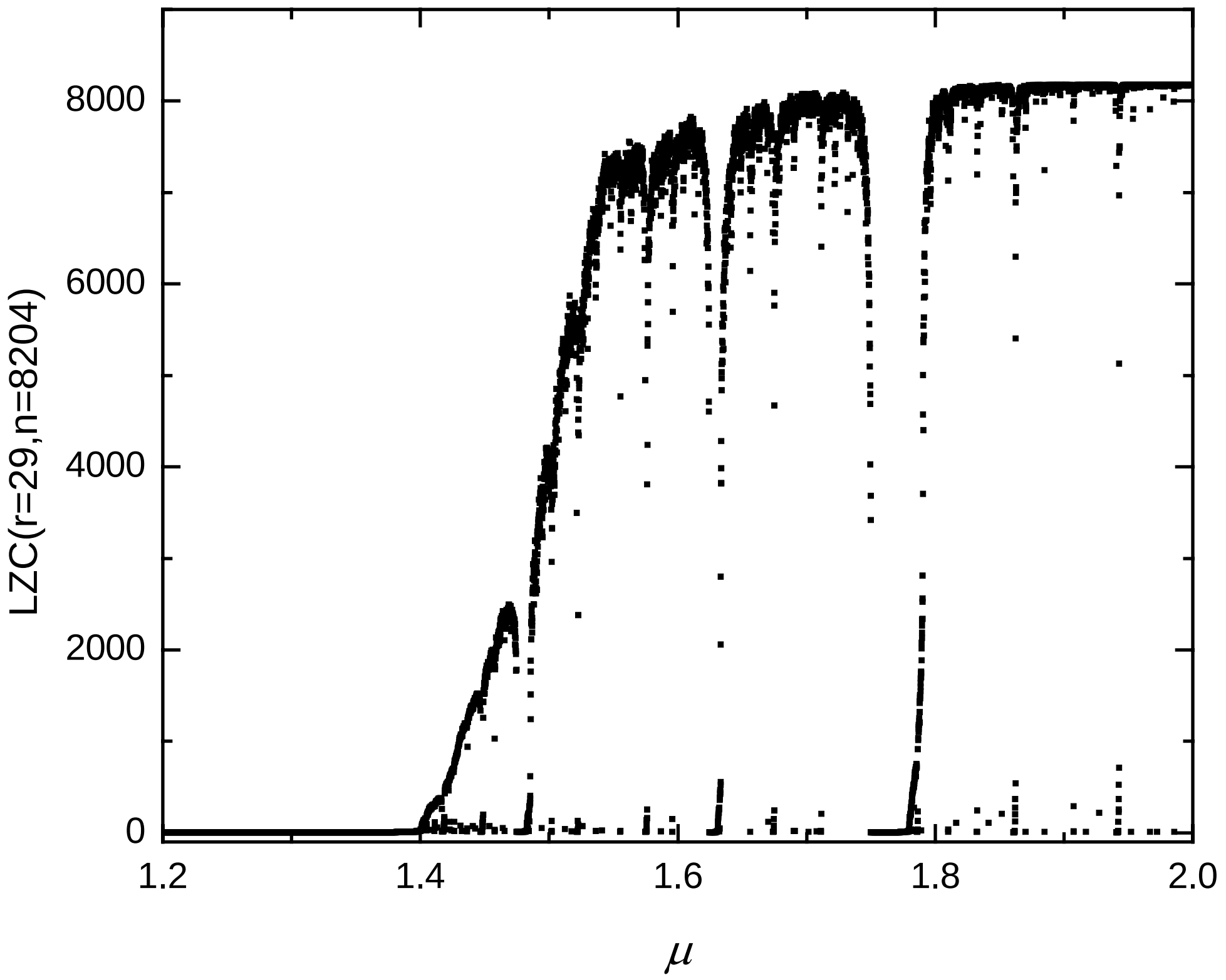

Fig. 7 Lempel-Ziv Complexity of Logistic map for $r = 29$

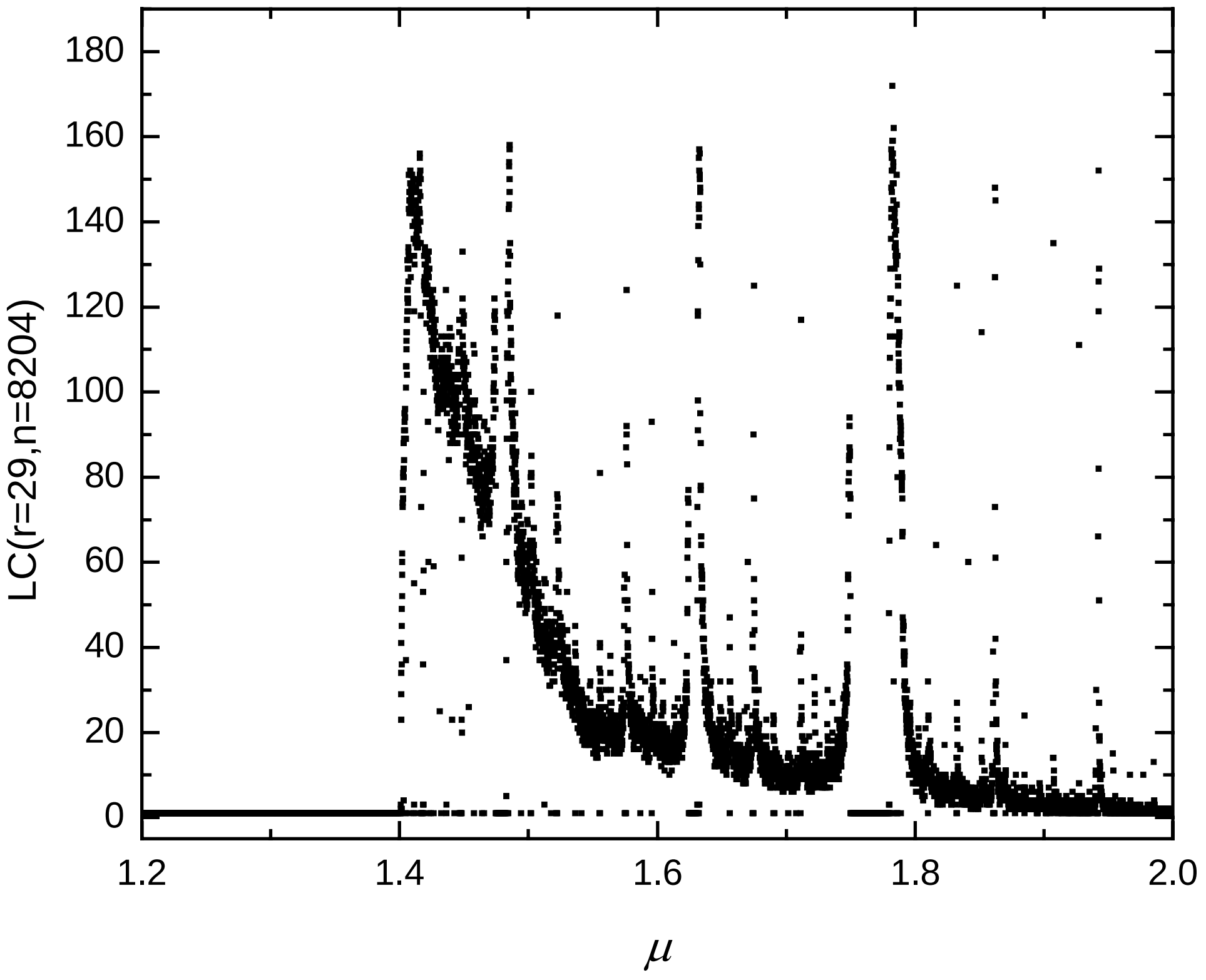

Fig. 8 *Lattice Complexity* of Logistic map for $r = 29$

From Fig. 7, we see that higher fine-graining orders generally indicate larger values of Lempel-Ziv Complexity for most sequences. But from Fig.8 we may easily find that, after the fine-graining order has enlarged to a certain level, *Lattice Complexity* takes the maximum roughly at the point where Logistic map goes from period-doubling route to chaos, and afterward there are other maximums on both sides of every periodic window. For the most random case, i.e. the full-blown chaos, *Lattice Complexity* equals 1, as the results of periodic sequences. That is to say, with $r$ augmenting, when more sequences are qualified as complex cases by Lempel-Ziv Complexity, such same sequences are regarded as simple cased by *Lattice Complexity*. The order making us to regard the completely chaotic sequence as the complex case will help to distinguish the meaning of complexity from that of randomness, while other orders will provide different choices for different applications.

# 5 Proofs and Discussions of Main Propositions

In addition to the properties mentioned in section 3, there are some other important propositions about the two measures and the fine-graining method.

Suppose we have a symbolic sequence $s$ of length $n$, and suppose there are $\alpha$ symbols in the basic alphabet $S$. We have the three propositions as follows:

**Proposition 1**: When $s$ is an $m$-periodic sequence ($m < n$),

$$LZC(m, n) = m, \text{ and } LC(m, n) = 1. \tag{11}$$

The proof of this proposition is obvious. Note that, without the assumption that every symbol represents only a single point in iterative maps, here the $m$-periodic sequence can have repeated symbols in one period.

**Proposition 2**: Let the same probability $\alpha^{-1}$ assign to every symbol in the alphabet $S$ to construct the sequence $s$, for $r \sim n$, we have

$$\lim_{n\to\infty}\lim_{r\to\infty}\Pr[LC(r,n) = 1] = 1 \tag{12}$$

and

$$\lim_{n\to\infty}\lim_{r\to\infty}\Pr[LZC(r,n) = n - r + 1] = 1. \tag{13}$$

Proof: From the conditions, we see that any sub-sequence of length $r$ has the possibility of $\alpha^{-r}$ to appear. Then the possibility of emerging two same $r$-sub-sequences is $\alpha^{-2r}$. Because in the sequence $s$ there are totally ($n - r + 1$) sub-sequences of length $r$, it must hold that

$$\Pr[LC(r,n) > 1] \le (n - r)\cdot\alpha^{-2r}.$$

Considering the condition $r \sim n$, we obtain

$$\lim_{n\to\infty}\lim_{r\to\infty}\Pr[LC(r,n) > 1] = 0$$

Because $LC(r,n) \ge 1$, (12) is valid. This illuminates that all the ($n - r + 1$) sub-sequences are distinct with probability 1, thus (13) is also valid.

Q.E.D

From proposition 2, we get two corollaries as follows:

**Corollary 1**: For $n = r + c$, where $c$ is a constant, we have

$$\lim_{r\to\infty}\Pr[LZC(r,n) = c + 1] = 1. \tag{14}$$

**Corollary 2**: When $n = c$ and $r \le c$, we have

$$\lim_{r \to c} \Pr[LZC(r,c) = c - r + 1] = 1 . \qquad (15)$$

Foregoing propositions show that by using *Lattice Complexity* we can regard periodic objects and random objects as simple. Only those having both regularity and randomness are hardly qualified as simple objects. The corollaries illustrate that for the random case, when $r$ is sufficiently large, a linear relationship is established between $LZC(r,n)$ and $r$. When $n$ is a constant, with $r$ increasing $LZC(r,n)$ will monotonically and linearly descend. Under the extreme condition $n = r$, the whole sequence is only one symbol that should be certainly regarded as the simplest no matter what measure is used.

Now we may define the *critical order* $r^*$ more explicitly: For a sequence $s$ of length $n$, the *critical order* $r^*$ is a special order such that for every $r \ge r^*$

$$LZC(r,n) = n - r + 1 \quad \text{or} \quad LC(r,n) = 1 .$$

A sequence of length $n = \alpha^k + k - 1$ with $\alpha$ symbols in its alphabet is called a de Bruijn sequence if all of its words of $k$ letters just exhaust all the $\alpha^k$ possibilities. It has already been known that the Lempel-Ziv Complexity of de Bruijn sequence is very high [7], but for *Lattice Complexity* we have a proposition as the following.

**Proposition 3**: If $s$ is a de Bruijn sequence of length $n = \alpha^k + k - 1 \quad (k \in N)$, we have

$$r^*(s) = k .$$

Proof: By the definitions of *Lattice Complexity* and de Bruijn sequence, we see that if $r = k$, the whole sequence is a single iterative sequence and $LC(k,n) = 1$.

Now we prove that $k$ is just the critical order. That will be true if $LC(r,n)$ remains as a constant for $r \ge k$ and takes a value larger than one for $r = k - 1$.

1) When $r > k$, all the sub-sequences of length $r$ in $s$ are not equal. Otherwise some sub-sequences of length $k$ will be equal, in contradiction to the definition of de Bruijn sequence. So $LC(r,n) = LC(k,n) = 1$. Namely, $LC(r,n)$ is stable for $r \ge k$.

2) When $r = k - 1$, there are totally $\alpha^{k-1}$ sub-sequences of length $(k-1)$, then there must be at least $\alpha$ sub-sequences not distinct in $s$. So the refined sequence $s^{k-1}$ is not a chaotic sequence. Nor is it a periodic sequence, otherwise the following letters of the equal sub-sequences are also equal so that some sub-sequences of length $k$ are also equal and contradictions occur. Hence, $LC(k-1,n) \ge \alpha > 1$, meaning that $LC(r,n)$ is unstable for .

Then we come to the conclusion that the critical order $r^*(s) = k$.

Q.E.D

From proposition 3 we see that the critical order of any binary de Bruijn sequence of length 8204 is 13. Comparing with the results of the time series made by Logistic map, it is easy to see that the number 13 is the minimum critical order among the all critical orders of non-periodic sequences of the same length. This provides a beneficial reference to the choice of fine-graining order in real applications. More explicitly, we do not need to search for a critical order by starting from 1.

Note that critical order is not always as obvious as shown in proposition 3, and it is not necessarily an integer. Since the value depends on the sequence length and the basic alphabet size, by changing the length and the size we should not exclude the possibility of obtaining a fraction value. We may take critical order as a special “fractal dimension” on the base of the coarse graining of the time series. Considering a multi-dimensional embedded space, when we divide each dimension into sub-intervals to make the partition according to a basic alphabet, the sub-intervals on all dimensions will constitute a multi-dimensional "box". Critical order is clearly the minimal order making any refined word in the sequence fall into each multi-dimension "box" not more than one time.

Actually, the definition of *Lattice Complexity* is built on a very strong supposition that each symbol only represents a recurrent point or a non-recurrent point. And the algorithm is based on a statistics rule that each symbol appears only once in a chaotic sequence. What will happen when we weaken the supposition? For example, we can define a chaotic sequence as a sequence such that each symbol appears $N$ ( ) times. For such a simple generalization of *Lattice Complexity*, almost all of the propositions in this section can find their generalized forms (only proposition 2 need some minor modifications), and Lempel-Ziv Complexity is obviously a special case for $N=0$. Bigger $N$ means smaller critical order.

On the other hand, because we choose a simple shift map on the symbolic sequence, associated to every step of the iterative map, with the fine-graining method we extract words of length $r$ as much as possible. If we take the symbolic sub-shift associated to more steps of the iterative map, there will be very interesting variations on the complexity measures. Although such variations are worth investigating carefully in a study of chaotic systems, here we do not make further discussions on them.

# 6 Conclusion

From our experiences and discussions, we see that *Lattice Complexity* can preserve the advantages of Lempel-Ziv Complexity and modify its defects. In complexity measurements, the fine-graining method of symbolic sequence acts like a "microscope" for observing structures of the sequence, and the fine-graining order acts like the focal length of this "microscope". With fine-graining method, we can measure complexity on different scales to reflect the hierarchy and the relativity of the complexity. With low order, *Lattice Complexity* and Lempel-Ziv Complexity are almost equal, whereas with high order they show contrary characteristics. This display a dialectical relationship between these two measures, like that of the two sides of a coin.

Our methods are applicable to every research field where symbolic sequences can be used to character the essential properties of the systems. Symbolic dynamics has already found its place on the areas from quantum mechanics [18] to traditional macrophysics [19]. For some areas where Lempel-Ziv Complexity or other complexity measure has already been applied, including analysis on electroencephalograph (EEG), electrocardiograph (ECG) and pseudo-random sequence[20] etc., our measures provide useful alternatives.

However, the most important is that this article publicizes and supports a very interesting idea about the complexity and the chaos. Under the condition of nonexistence of absolute universal complexity character, most sequences can be regarded as both complex and simple ones. But with a proper fine-graining order, there are some workable measures, e.g. *Lattice Complexity*, making a few sequences (particularly the sequence near the edge of the chaotic area) show objective differences in complexity with others. This may represent a certain essential character of natural complex things—emerging "at the edge of order and chaos" [21].

# Acknowledgements

This work has supported in part by National Natural Science Foundation of China (Grant No: 30170267) and National Pre-project for Base Research (No: 2002CCA01800).

---

[1]. B. L. Hao. 2001 Physics **30** 466 (in Chinese)
[2]. A.N. Kolmogorov, Info. Trans. 1965; **1**: 3
[3]. Remo. Badii and Antonio. Politi, 1997 Complexity: hierarchical structures and scaling in physics (Cambridge: Cambridge University Press) 233~235
[4]. F. Kaspar, H.G. Schuster, 1987 Physical Rev. **A 36** 842
[5]. Ling-Yun Li, Qin-Ye Tong, 2000 Acta Electr. Sin. **28** 97 (in Chinese)
[6]. M. D. Li, H. Zhang, Q. Y. Tong, 2003 Space Medicine and Medical Engineering **16** 215 (in Chinese)
[7]. A. Lempel, Ziv, 1976 IEEE Trans. Inform. Theor. **IT-22** 75
[8]. H. M. Xie, 1994 Complexity and Dynamical System (Shanghai: Shanghai Scientific and Technological Education Publishing House) (in Chinese)
[9]. P. Grassberger, 1986 Inter. J Theor. Phys. **25** 907
[10]. P. W. Anderson, 1991 Physics Today. **7** 9
[11]. W. M. Zheng, B. L. Hao, 1994 Applied Symbolic Dynamics (Shanghai: Shanghai Scientific and Technological Education Publishing House) (in Chinese)
[12]. Z. J. Zhang, S. G. Chen, 1989 Acta Phys. Sin.**38** 1 (in Chinese)
[13]. Z. L. Zhou, 1997 Symbolic Dynamics (Shanghai: Shanghai Scientific and Technological Education Publishing House) (in Chinese)
[14]. R. L. Devaney, 1989 An Introduction to Chaotic Dynamical Systems. Second Edition. ( Addison-Wesley, Redwood City, Calif.)
[15]. R. Wackerbauer et al, 1994 Chaos, Solitons & Fractals. **4** 133
[16]. X. Meng et al, 2000 Acta Biophysica Sin. **16** 701 (in Chinese)
[17]. E. H. Shen et al, 2000 Acta Biophysica Sin. **16** 707 (in Chinese)
[18]. E. B. Boqomolny, 1992 Nonlinearity **5** 805
[19]. P. Walters, Editor, 1992 Symbolic Dynamics and Its Applications. Contemporary Mathematics Vol.135. (Providence, R1: Oxford: American Mathematical Society)
[20]. J. P. Cai, Z. Li, W. T. Song, 2003 Acta Phys. Sin. **52** 1873 (in Chinese)
[21]. M. M. Waldrop. Complexity: 1992 The Emerging Science at the Edge of Order and Chaos (NewYork: Simon and Shuster)